\documentclass[aps, prd, preprint, 12pt, superscriptaddress]{revtex4-2}

\usepackage[utf8]{inputenc}
\usepackage[normalem]{ulem}
\usepackage{amsmath} \usepackage{amsfonts} \usepackage{amssymb}
\usepackage{wasysym}
\usepackage[dvipsnames]{xcolor}
\usepackage{graphicx}
\usepackage{multirow}
\usepackage{here}
\usepackage{epsfig}
\usepackage{epstopdf}
\usepackage{soul}
\usepackage{hyperref}
\usepackage{natbib}
\usepackage{graphicx}
\usepackage{verbatim}
\usepackage{mathtools}
\usepackage{slashed}
\usepackage{cancel}
\usepackage{tikz} 
\usepackage{caption}

\newcommand{\fn}[2]{\mathinner{#1\mathopen{\left(#2\right)}}}
\newcommand{\eq}[1]{Eq.~(\ref{#1})}
\newcommand{\eqs}[2]{Eqs.~(\ref{#1}) and (\ref{#2})}
\newcommand{\eqss}[3]{Eqs.~(\ref{#1}), (\ref{#2}) and (\ref{#3})}

\newcommand{\Rr}{\mathcal{R}_{\delta\rho_\mathrm{r}}}

\newcommand{\dr}{\delta \rho_\mathrm{r}}

\newcommand{\drhorr}{\delta\rho_{\mathrm{r}_\mathcal{R}}}
\newcommand{\dotdrhorr}{\delta\dot{\rho}_{\mathrm{r}_\mathcal{R}}}
\newcommand{\ddotdrhorr}{\delta\ddot{\rho}_{\mathrm{r}_\mathcal{R}}}

\begin{document}

\title{The curvature perturbation generated by thermal fluctuations during thermal inflation}

\author{Jeong-Myeong Bae}
\email{bjmhk2@snu.ac.kr}
\affiliation{Center for Theoretical Physics, Department of Physics and Astronomy, Seoul National University, Seoul 08826, Korea}
\author{Hammam Raihan Mohammad}
\affiliation{DataHen Canada Inc., 30 Wellington St,
Toronto, Canada}
\author{Ewan D. Stewart}
\affiliation{Department of Physics, Izmir Institute of Technology, Gulbace, Urla 35430, Izmir, Turkiye}
\author{Heeseung Zoe}
\email{heeseungzoe@iyte.edu.tr}
\affiliation{Department of Physics, Izmir Institute of Technology, Gulbace, Urla 35430, Izmir, Turkiye}

\date{\today}

\begin{abstract}
During thermal inflation, the temperature determines the number of e-folds of expansion of the universe and so  thermal fluctuations are magnified into curvature perturbations.
We use classical thermodynamics to calculate the subhorizon thermal fluctuations and trace their evolution into  superhorizon temperature perturbations. 
We convert the temperature perturbations into curvature perturbations using the $\delta N$-formalism, or equivalently the junction condition of curvature perturbations at the end of thermal inflation, denoted by subscript c, and show that the late-time power spectrum is $P_\mathcal{R} = \frac{15}{4\pi^4} \frac{H^3_\mathrm{c}}{g_* T^3_\mathrm{c}} \frac{k^3}{k^3_\mathrm{c}}$. 
\end{abstract}

%\pacs{98.80.Bp, 98.80.Cq}

\maketitle

\newpage

\section{Introduction}

In the case of primordial inflation, subhorizon {\it vacuum} fluctuations are magnified into  superhorizon curvature perturbations. 
Well after primordial inflation, moduli, expected in string theory, are generated and immediately dominate the energy density of the universe causing serious problems \cite{Coughlan:1983ci, deCarlos:1993wie, Banks:1993en}.  
Thermal inflation, a brief inflation driven by the potential of a flaton, an unstable flat direction, when it is held at the origin by thermal effects, can dilute the moduli to a safe abundance solving this moduli problem \cite{Lyth:1995hj, Lyth:1995ka}.     
Small-scale primordial perturbations that enter the horizon during the moduli domination before thermal inflation and leave during thermal inflation are suppressed \cite{Hong:2015oqa, Bae:2022gkv, Cho:2017zkj, Hong:2017knn, Leo:2018kxp}.

However, thermal inflation also generates its own perturbations. 
During thermal inflation, suhorizon {\it thermal} fluctuations are magnified into superhorizon temperature perturbations. 
At the end of thermal inflation, which occurs when the temperature drops below a critical temperature, the temperature perturbations are converted into  superhorizon curvature perturbations. 
These perturbations have a spectral index of $n = 4$ with a peak at the horizon size at the end of thermal inflation, as estimated in Appendix A.3 of \cite{Kadota:2003fs}. 
In this note, we calculate these perturbations precisely.

One of the motivations of this work is to correct the misconceptions of \cite{Dimopoulos:2019wew, Lewicki:2021xku, Bastero-Gil:2023sub}, in which the authors simply assumed {\it flaton} fluctuations generate the curvature perturbations as the inflaton fluctuations do in primordial inflation. 
However, the flaton is held at the origin during thermal inflation and so the flaton fluctuations are irrelevant.
Instead, it is {\it thermal} fluctuations that are magnified into curvature perturbations as described above and in more detail in Section \ref{sec:tf} and \ref{sec:cp}.

In Section \ref{sec:tf}, we calculate the subhorizon thermal fluctuations using classical thermodynamics and evolve them into superhorizon temperature perturbations. 
In Section \ref{sec:cp}, we use the $\delta N$ formalism \cite{Sasaki:1995aw}, or equivalently the junction condition at the end of thermal inflation \cite{Hong:2015oqa}, to convert the temperature perturbations into  curvature perturbations and determine their late-time power spectrum. 
In Section \ref{sec:dis}, we discuss the implications of our results.

\section{Subhorizon Thermal Fluctuations to Superhorizon Temperature Fluctuations }\label{sec:tf}

During thermal inflation, the temperature is much greater than the Hubble parameter, $T \gg H$, and,  for modes well inside the horizon, the physical wave number is much greater than the Hubble parameter,  $q \gg H$, so thermalization is faster than the Hubble expansion and the subhorizon temperature fluctuations can be described by classical thermodynamics in flat spacetime \cite{Landau:1980mil}.
The temperature fluctuations in a volume $V$ are then 
\begin{equation}
\Delta T = \frac{1}{V} \int_V d^3x \fn{\delta T}{x}
\label{eq:DTdef}
\end{equation}
where 
\begin{equation}
\langle\Delta T^2\rangle = \frac{T^2}{C_V}
\label{eq:DTCV}
\end{equation}
and the heat capacity of a thermal bath of radiation is
\begin{eqnarray} 
C_V = \left( \frac{\partial U}{\partial T} \right)_V = \frac{2\pi^2}{15} g_* T^3 V~.
\label{eq:CV}
\end{eqnarray}
The corresponding power spectrum of temperature fluctuations
\begin{eqnarray} 
\langle \fn{\delta T}{q} \fn{\delta T^*}{q'} \rangle = \frac{2\pi^2}{q^3} \fn{P_{\delta T}}{q} \fn{\delta^3}{q-q'}
\label{eq:Pqdef}
\end{eqnarray}
is given by 
\begin{equation}
\fn{P_{\delta T}}{q} = \frac{15}{4\pi^4} \frac{q^3}{g_* T}~.
\label{Pq}
\end{equation}
which is derived in detail in Appendix \ref{app:dimension}.

Applying this to a subhorizon scale at an intial time $t=t_\mathrm{i}$ during thermal inflation, \eq{Pq} becomes 
\begin{eqnarray} \label{initial_1}
\fn{P_{\delta T}}{k} = \frac{15}{4\pi^4} \frac{1}{g_* T_\mathrm{i}} \frac{k^3}{a^3_\mathrm{i}}
\end{eqnarray}
which is derived in detail in Appendix \ref{app:conversion}, which then evolves as the universe expands during thermal inflation. 
Thermal equilibrium is preserved by the expansion, and so the form of subhorizon thermal fluctuations is preserved as it evolves to superhorizon scales
\begin{eqnarray} 
\fn{P_{\delta T}}{k} = \frac{15}{4\pi^4} \frac{1}{g_* T} \frac{k^3}{a^3}
\label{PdT}
\end{eqnarray}
which is derived in detail in Appendix \ref{app:tp}.

\section{Temperature Perturbations to Curvature Perturbations}\label{sec:cp}

The $\delta N$ formalism \cite{Sasaki:1995aw} gives the change in the superhorizon curvature perturbations between two times $t_1$ and $t_2$ in terms of the number of e-folds 
\begin{eqnarray} 
\fn{\mathcal{R}}{t_2} - \fn{\mathcal{R}}{t_1} = \delta N~.
\end{eqnarray} 
Taking a flat hypersurface at $t_1$, this reduces to 
\begin{eqnarray}
\fn{\mathcal{R}}{t_2} = \delta N
\end{eqnarray} 

In the usual primordial slow-roll inflation with a single component inflaton, the inflaton $\varphi$ determines the number of e-folds,  $N = \fn{N}{\varphi}$, and so 
\begin{eqnarray}
\mathcal{R} = \frac{\partial N}{\partial \varphi} \delta\varphi
\end{eqnarray} 
while the slow-roll condition leads to
\begin{equation}
\frac{\partial N}{\partial \varphi} = -\frac{H}{\dot{\varphi}}.
\end{equation}

In thermal inflation,  inflation ends when the temperature drops below a critical temperature. Thus, the temperature determines the number of e-folds, $N = \fn{N}{T}$, and so
\begin{eqnarray} 
\mathcal{R} = \frac{\partial N}{\partial T} \delta T
\label{ccp}
\end{eqnarray}
where the relation between temperature and spatial expansion gives
\begin{equation}
\frac{\partial N}{\partial T} = \frac{1}{T}.
\label{dNdT}
\end{equation}

We can also derive \eqs{ccp}{dNdT} by matching the curvature perturbations on the constant radiation energy density hypersurface just before the end of thermal inflation
\begin{equation}
\Rr = \mathcal{R} - \frac{\dr}{\dot{\rho}_\mathrm{r}}H
\end{equation}
to the curvature perturbation on the constant density hypersurface just after the end of thermal inflation
\begin{equation}
\mathcal{R}_{\delta \rho} = \mathcal{R} - \frac{\delta \rho}{\dot{\rho}}H
\end{equation}
giving
\begin{equation}
\fn{\Rr}{t_\mathrm{c}^{-}} = \fn{\mathcal{R}_{\delta \rho}}{t_\mathrm{c}^+}
\end{equation}
where subscript c denotes the end of thermal inflation \cite{Hong:2015oqa}.
During thermal inflation, the curvature perturbations on constant radiation energy density hypersurfaces are related to the temperature perturbations on flat hypersurfaces   
\begin{equation}
\Rr =  -\frac{H}{\dot{\rho}_\mathrm{r}} \drhorr = \frac{\delta T}{T}
\label{Tfluc}
\end{equation}
in agreement with \eqs{ccp}{dNdT}.

Therefore, the power spectrum of the late-time superhorizon curvature perturbations on constant energy density, or equivalently comoving, hypersurfaces is 
\begin{eqnarray}
P_{\mathcal{R}} = \left(\frac{\partial N}{\partial T}\right)^2 P_{\delta T}
\label{Prform}
\end{eqnarray}
leading to 
\begin{eqnarray}
P_{\mathcal{R}} = \frac{15}{4\pi^4} \frac{H^3_\mathrm{c}}{g_* T^3_\mathrm{c}} \frac{k^3}{k^3_\mathrm{c}} 
\label{eq:ps}
\end{eqnarray}
where $k_\mathrm{c} = a_\mathrm{c} H_\mathrm{c}$ is the comoving horizon scale at the end of thermal inflation.
\eq{eq:ps} is valid for modes that leave the horizon during thermal inflation, $k_\mathrm{b} \ll k \ll k_\mathrm{c}$, where subscript b denotes the beginning of thermal inflation.

\section{Discussion}\label{sec:dis}

In this note, we calculate the curvature perturbations generated by thermal fluctuations during thermal inflation.  
We use classical thermodynamics in flat spacetime to calculate the thermal fluctuations well inside the horizon and evolve them  into superhorizon temperature perturbations obtaining the temperature power spectrum of \eq{PdT}. 
We apply  the $\delta N$ formalism, or equivalently the junction condition  at the end of thermal inflation, to \eq{PdT} and derive the late-time curvature perturbations of \eq{eq:ps}. 
In \cite{Dimopoulos:2019wew, Lewicki:2021xku, Bastero-Gil:2023sub}, the authors incorrectly considered flaton fluctuations, instead of thermal fluctuations, to calculate the curvature perturbations and so arrived at incorrect results. 
Our power spectrum can also be  potentially probed by primordial black holes \cite{Dimopoulos:2019wew} and induced gravitational waves \cite{Lewicki:2021xku}.

\acknowledgements
This work is supported by T\"{U}B\.{I}TAK-ARDEB-1001 program under project 123F257.
JB is supported by Grant Korea NRF2019R1C1C1010050 and RS-2024-00342093. The authors thank Junghwan Lee, Enes Zeybek, Hyukjung Kim, Tae Hyun Jung and Aykut Aktan for helpful discussions. 
HZ thanks Kiwoon Choi, Sanghyeon Chang, Jai-chan Hwang, Sang Hui Im, Seokhoon Yun and IBS-CTPU for their hospitality.

\appendix

\section{Thermal fluctuation power spectrum}\label{app:dimension}

Using \eqs{eq:DTdef}{eq:Pqdef}, the dispersion of thermal fluctuations in a sphere of radius $R$ is 
\begin{eqnarray}
\langle \Delta T^2  \rangle 
= \frac{1}{V^2} \int_{V} d^3x \int_{V} d^3 x' \langle \fn{\delta T}{x} \fn{\delta T^*}{x'} \rangle  = \frac{R^7}{V^2} \int dq \fn{W}{qR} \fn{P_{\delta T}}{q}
\label{eq:DTW}
\end{eqnarray}
where 
\begin{equation}
\fn{W}{x} = \frac{ 16 \pi^2 }{x^7} \left( \sin x - x \cos x \right)^2~.
\end{equation}
Using the ansatz
\begin{equation}
\fn{P_{\delta T}}{q} = A q^\alpha ~,
\end{equation}
matching to \eqs{eq:DTCV}{eq:CV} gives
\begin{equation}
A = \frac{15}{4 \pi^4 g_* T}, \quad \alpha = 3.
\end{equation}
Therefore, the power spectrum of thermal fluctuations is
\begin{equation}
\fn{P_{\delta T}}{q} = \frac{15}{4\pi^4} \frac{q^3}{g_* T}~.
\label{eq:PdTapp}
\end{equation}

\section{Conversion between physical and comoving thermal fluctuations}\label{app:conversion}

We define the physical Fourier component of 
the thermal fluctuations by 
\begin{equation}
\delta T 
= \frac{1}{\left( 2\pi \right)^{3/2}} 
\int d^3q \fn{\delta T}{q }e^{i q \cdot x_p},
\end{equation} 
and the comoving Fourier component by
\begin{equation}
\delta T 
= \frac{1}{\left( 2\pi \right)^{3/2}} 
\int d^3k  \fn{\delta T}{k }e^{i k \cdot x_c}.
\end{equation}
\eq{eq:Pqdef} is 
\begin{eqnarray} 
\langle \fn{\delta T}{q} \fn{\delta T^*}{q'} \rangle 
= \frac{2\pi^2}{q^3}  \fn{P_{\delta T}}{q} \fn{\delta^3}{q-q'}.  
\label{eq:Tq1}
\end{eqnarray}
and similarly 
\begin{eqnarray} 
\langle \fn{\delta T}{k} \fn{\delta T^*}{k'} \rangle  = \frac{2\pi^2}{k^3}  \fn{P_{\delta T}}{k} \fn{\delta^3}{k-k'}.
\label{eq:Tq2}
\end{eqnarray}
Due to the different Fourier measures, using   $q=k/a$, we get 
 \begin{equation}
\fn{\delta T}{q} = a^3\fn{\delta T}{k }.
\label{eq:dTq}
\end{equation}
Using 
\begin{equation}
\fn{\delta^3}{q-q'} = a^3 \fn{\delta^3}{k-k'}~,
\end{equation}
\eqss{eq:Tq1}{eq:Tq2}{eq:dTq} give
\begin{equation}
\fn{P_{\delta T}}{k} = \fn{P_{\delta T}}{q}
\end{equation}
implying that \eq{Pq} is simply converted to \eq{initial_1}.

\section{Evolution of the temperature perturbation power spectrum}\label{app:tp}

To model the evolution of the subhorizon thermal fluctuations for $t > t_\mathrm{i}$, we consider a simple system of vacuum energy and radiation. 
For modes well inside the horizon,  standard perturbation theory \cite{Kodama:1984ziu} gives 
\begin{eqnarray}
\ddotdrhorr + 9H \dotdrhorr + \left( 20H^2 + \frac{q^2}{3}\right)\drhorr = 0 ~.
\end{eqnarray}
which has solution
\begin{eqnarray}
\fn{\drhorr}{k,t} & = & 4 \rho_\mathrm{r}\fn{C_1}{k}  \exp\left( i \int \frac{k}{\sqrt{3}\,a} dt \right)  + 4 \rho_\mathrm{r}\fn{C_2}{k}  \exp\left( -i \int \frac{k}{\sqrt{3}\,a} dt \right)~.
\label{eq:rhorrreq}
\end{eqnarray}
Using
\begin{equation}
\frac{\delta T}{T} = \frac{\drhorr}{4\rho_\mathrm{r}}~,
\end{equation} 
\eq{eq:rhorrreq} gives the temperature perturbation
\begin{eqnarray}
\fn{\delta T}{k,t} = T \fn{C_1}{k}\exp\left( i \int \frac{k}{\sqrt{3}\,a} dt \right)  + T \fn{C_2}{k} \exp\left( -i \int \frac{k}{\sqrt{3 \,}a} dt \right)~.
\end{eqnarray}
Taking the ensemble average over random phases
\begin{eqnarray} 
\left\langle  \fn{C_i}{k}\fn{C^*_j}{k'} \right\rangle
= \frac{2\pi^2}{k^3} \fn{P_i}{k} \delta_{ij}\fn{\delta^3}{k-k'} ~,
\end{eqnarray}
we get 
\begin{eqnarray} 
\langle \fn{\delta T}{k,t} \fn{\delta T^*}{k',t} \rangle   
= T^2 
\left( \frac{2\pi^2}{k^3} \right)
\left[\fn{P_1}{k} + \fn{P_2}{k}
 \right] \fn{\delta^3}{k-k'}~.
\label{eq:2ptPP}
\end{eqnarray}
By combining \eq{eq:2ptPP} with \eq{eq:Tq2}, we conclude 
\begin{eqnarray}
\fn{P_{\delta T}}{k} \propto \frac{1}{a^2}~.
\label{eq:PT}
\end{eqnarray}

\bibliographystyle{apsrev4-2}
\bibliography{tcpbib}

\end{document}